# Integrated Quantum Controlled-NOT Gate Based on Dielectric-Loaded Surface Plasmon Polariton Waveguide

S. M. Wang[1,*,†], Q. Q. Cheng[2,*], Y. X. Gong[3], P. Xu[1], L. Li[1], T. Li[2,†], and S. N. Zhu[1,†]

[1]*National Laboratory of Solid State Microstructures and School of Physics, Nanjing University, Nanjing 210093, China.*
[2]*National Laboratory of Solid State Microstructures and College of Engineering and Applied Sciences, Nanjing University, Nanjing 210093, China.*
[3]*Department of Physics, Southeast University, Nanjing 211189, China.*

*These authors contributed equally to this work.
[†]Corresponding authors. Email: wangshuming@nju.edu.cn (S.M.W.); taoli@nju.edu.cn (T.L.); zhusn@nju.edu.cn (S.N.Z.).

**Abstract:** It has been proved that surface plasmon polariton (SPP) can well conserve and transmit the quantum nature of entangled photons. Therefore, further utilization and manipulation of such quantum nature of SPP in a plasmonic chip will be the next task for scientists in this field. In quantum logic circuits, the controlled-NOT (CNOT) gate is the key building block. Here, we implement the first plasmonic quantum CNOT gate with several-micrometer footprint by utilizing a single polarization-dependent beam-splitter (PDBS) fabricated on the dielectric-loaded SPP waveguide (DLSPPW). The quantum logic function of the CNOT gate is characterized by the truth table with an average fidelity of *79.6 ± 7.7%*. Its entangling ability to transform a separable state into an entangled state is demonstrated with the visibilities of *84.4 ± 9.0%* and *78.8 ± 10.1%* for non-orthogonal bases. The DLSPPW based CNOT gate is considered to have good integratability and scalability, which will pave a new way for quantum information science.

Quantum information science has recently attracted worldwide attention since its dramatic speedup of the performance in storing, processing and transmitting information by harnessing quantum effects. Among diverse quantum systems, photons are good candidates for realizing various quantum information tasks. However, the implementations of quantum optical circuits based on traditional bulk Optics were still limited by the size, purity and stability, until the integrated photonic circuits were introduced to process quantum information [1]. Integrated photonic circuits based on silica-on-silicon or femtosecond-laser-writing waveguides have proved their power and advantage in realizing basic quantum logic gates [2-5] and some crucial quantum algorithms [6, 7]. In quantum integration community, Quantum Plasmonics has been a fast advancing field in the last decade, wherein the SPP works as a good carrier of quantum information as photon. The quantum properties of photons can be well conserved through exciting such macroscopic collective resonances of electrons at the metal surface [8]. For the compact characteristic, the SPP beam-splitters own a much smaller size than conventional integrated photonic circuits based on dielectric waveguides, which also present good quality in quantum interference [9-11]. Therefore, the plasmonic circuit becomes a promising choice for realizing integrated quantum building blocks. While the conventional plasmonic chip based on simple metal surface is polarization-limited, which only can steer TM polarized wave. To overcome this problem, the dielectric-loaded SPP waveguide

(DLSPPW) is introduced, which allows the existence of TE mode wave. And this approach offers a good opportunity to separately steer these two perpendicularly polarized waves, due to their discrepant characters in wave vector and field distribution. Therefore, the polarization-encoded information processing is possible to be realized in the DLSPPW system. Besides, the DLSPPW devices also enable more control degrees for the optical nonlinearity, thermal and electro-optical properties of the loaded dielectrics [*12-15*].

Here, we implement a polarization-encoded quantum CNOT gate based on the DLSPPW system, which is the pivotal element able to form any kind of quantum computing circuits combined with different single-qubit gates [*16*]. The widely used linear optical quantum CNOT gate [*17*], or the equivalent quantum controlled-phase (C-phase) gate [*18*], has been realized in the bulk optical system [*19*] and waveguide system [*1*], employing path interference with several beam splitters. By utilizing polarization-encoded qubits, the CNOT gate was simplified into three PDBSs with a reduced total size and increased stability for the absence of phase-sensitive interference [*20-22*]. Very recently, this simplified CNOT gate has been realized via femtosecond-laser-writing waveguides with a millimeter size [*4*]. In this paper, with the capability of steering both TM and TE waves, we realized a PDBS in the DLSPPW system. By using such PDBS, we first implement the polarization-encoded quantum CNOT gate with only a several micrometer footprint. Moreover, since the specially designed output slits can manage the function of two attenuating PDBSs in the previously proposed version, we further simplifies the architecture of the CNOT gate into only single PDBS based on DLSPPW [*4, 20-22*].

A schematic representation of working mechanism of the proposed PDBS base on DLSPPW system is shown in Fig. 1 A. The DLSPPW is composed of a *300nm* silver film and a silica film covered on it with the thickness of *250nm*, which only allows the existence of the lowest TM mode wave (SPP) and TE mode wave in the system [*23*]. The input gratings milled on the silver film by focused ion beam etching (FIB, Helios nanolab 600i, FEI) convert spatial light into SPP at the input area. The SPP together with the higher TM mode waves can be excited by spatial *p*-polarized light with the electric field normal to grating groove, and the TE mode waves correspond to the *s*-polarized light with the electric field along the grating groove. The SPP propagates along the metal surface and meets a *45 °* oblique grating milled on the silver surface, which works as a SPP reflector whose transmittance and reflection can be flexibly controlled via carefully choosing its parameters. For the CNOT gate, the transmittance/reflection (T/R) ratio is required to be *1:2* [*20*]. The exact values of transmittance and reflection of SPP in our gate are respectively *1/3ξ* and *2/3ξ*, where *ξ* is related to the total loss of the system, including the converting loss, propagating loss, *etc*. For extraction and measurement of the quantum information in the gate, we use output slits to convert SPP again into spatial *p*-polarized light at the back side of the sample, which is finally collected by an oil objective. A scanning electron microscopy (SEM) image of the configuration of the sample at the surface of silver film is presented in Fig. 1 B. The CCD images of the output areas with the *785nm* laser illuminating each input grating are plotted in Fig. 1 C and D. The measured T/R ratios are respectively *1:1.9* and *1:2.15*, which are close to the ideal requirement ratio *1:2* [*24*]. Here, *ξ* can be estimated to be about *1%* from the experimental results. The difference between these T/R ratios of these two input gratings may be resulted from the fabrication defection of FIB etching.

On the other hand, the spatial *s*-polarized light can also be converted into the lowest TE mode wave of the DLSPPW at the same input gratings for SPP. However, the reflector of SPP has hardly any influence on TE mode wave, since the wavelengths of the TE mode wave and SPP are different. This leads to a nearly unity transmittance of TE mode wave at the reflector, which satisfies the required perfect transmission for the CNOT gate [20]. Moreover, the output slits can not only convert TE mode wave into spatial *s*-polarized light for collection but also can tune the transmittance to *1/3ξ* by exactly controlling of their parameters, in order to balance the contributions of TE mode wave and SPP in the gate. Therefore, the two additional PDBSs for TE mode wave attenuation in previously proposed model can be removed [4, 20-22], and the architecture of the polarization-encoded CNOT gate is further simplified into only a single PDBS based on DLSPPW system. Thus, the ultra-high integration of quantum circuit can be proposed. The CCD image at the output areas with *s*-polarized light input is shown in Fig. 1 E. The transmittances are respectively *37%ξ* and *35%ξ* with input from the right and left grating. Nearly zero reflection of TE mode wave can be observed in both cases. Since the results at both input gratings are quite similar, we only present the case with left grating input for simplicity. The *45 °*alignment of the whole device is exploited for the convenience of separate collection of the photons at two outputs.

We now demonstrate how to use this DLSPPW based gate to realize the functions of the quantum CNOT gate. We define the logical computational basis of the gate as $|0\rangle_c \equiv |s\rangle_c$, $|1\rangle_c \equiv |p\rangle_c$ for the control qubit, and $|0\rangle_t \equiv |D\rangle_t = (|s\rangle_t + |p\rangle_t)/\sqrt{2}$, $|1\rangle_t \equiv |A\rangle_t = (|s\rangle_t - |p\rangle_t)/\sqrt{2}$ for the target qubit. Under post-selection, *i.e.*, a successful twofold coincidence measurement on the two output ports, the CNOT gate succeeds with a probability of *1/9* [17, 18].

A femtosecond laser with the wavelength of *392.5nm* (the second harmonic generation of the *785nm* femtosecond laser) is used to pump a type-II BBO crystal to generate photon pairs with orthogonal polarizations [25]. We first investigate the quality of the photon pair source by using the two-photon Hong-Ou-Mandel (HOM) interference [26]. The HOM interference of the photon pair source is shown in Fig. 3 A. An obvious dip with a high visibility $V = (C_{max} - C_{min})/C_{max} = 95.2 \pm 0.7\%$, shows its good quality. The photon pairs, separated by a polarizing beam splitter (PBS), are collected by single mode fibers and injected to the DLSPPW based gate to excite the SPP with their polarizations both rotated into *p*-polarization. The best interference at the DLSPPW-based PDBS can be obtained through careful control the temporal overlap with the delay line. A HOM interference of the SPPs is plotted in Fig. 3 B, with a visibility of *72.4±3.1%*, which is much satisfying considering the theoretical value of *80%* for an ideal *1:2* beam splitter. This also proves the good quality of SPP for being a quantum information carrier.

To characterize the operation of this gate, we first measured the output of the gate for each of the four possible logical computational basis input states: $|00\rangle_{ct}$, $|01\rangle_{ct}$, $|10\rangle_{ct}$, and $|11\rangle_{ct}$. The truth table of the gate is presented in Fig. 4 A, where the probabilities are obtained by normalization with the sum of all coincidence counts obtained for the respective input state. The probabilities of *87.5±8.1%*, *86.9±8.0%* for the input states $|00\rangle_{ct}$, $|01\rangle_{ct}$ are higher than those of *72.1±7.5%*, *71.9±7.3%* for $|10\rangle_{ct}$, $|11\rangle_{ct}$, because the non-classical interference relying on the overlap on the PDBS is required for the two latter input states while is not required for the two former input states. The average fidelity of the logical computational basis can be obtained as $F = 79.6 \pm 7.7\%$. The results show that our DLSPPW based gate can well present the logic function of a CNOT gate.

The next step is to demonstrate that the gate can be used as an entangling gate, which produces a two-qubit entangled output state from a separable input state by measuring conditional fringe visibilities [19]. Figure 4 B shows the coincidence as a function of the HWP setting in the target polarization controller. The two curves are for the control output state set to $|0\rangle_c$ and $(|0\rangle_c + |1\rangle_c)/\sqrt{2}$, respectively. In both cases, the input state is $(|0\rangle_c - |1\rangle_c)|1\rangle_t / \sqrt{2}$, which may have the result in the output state as the maximally entangled Bell state $|\psi^-\rangle = (|01\rangle - |10\rangle)/\sqrt{2}$ after the CNOT gate. The visibilities obtained by $V = (C_{max} - C_{min})/(C_{max} + C_{min})$, are $84.4 \pm 9.0\%$ and $78.8 \pm 10.1\%$ for the two fitted curves, respectively, which are both above $78\%$ and hence the output states can both violate Bell inequalities [27]. Consequently, the high visibilities show a typical signature of entanglement creation of the gate.

It should be mentioned that for extraction and measurement of quantum information in the integrated gate, the output slits of the PDBS are required to convert both SPP and TE mode wave into spatial light. In fact, such conversion is not necessary for the integratability of quantum plasmonic devices in an actual chip. Without conversion to spatial light, the total loss of the system will be dramatically reduced, which will greatly improve the performance of the CNOT gate. And the attenuation function for TE mode wave of these slits can easily be managed by some additional small structures, such as gratings milled in the loaded dielectric. By utilizing nano-fabrication techniques, the plasmonic function elements can conveniently be integrated in a chip to realize diverse logic functions or algorithms within a rather small size [28]. For the wavelength of *785nm* in vacuum, the propagating length of light in DLSPPW is more than *100μm* theoretically, which is able to contain many plasmonic function elements. The compact characteristic of the gate also can overcome the decoherence effect resulting from the different velocities of two polarized waves, because the phase mismatch proportional to the total size of gate is much smaller than the coherence length of the photon pair source in our experiment [29]. And the coherence length can be further increased by using a narrower interference filter behind the photon pair source, which will meet the requirement for the scalability of the quantum plasmonic chip. In addition, since the plasmonic system provides opportunities to efficiently extract SPPs from integrated single-photon emitters [30, 31], combined with recently advancing technologies in integrated superconducting single-photon detectors [9], the total loss and the total size will be further reduced and the entire quantum system, including sources, logic gates and detectors, will be really integrated in a plasmonic chip.

Here, we have for the first time implemented the polarization-encoded integrated quantum CNOT gate based on DLSPPW in only several micrometers size. Our results directly prove that the SPPs are capable to be used to manipulate the quantum information. A specially designed PDBS and output slits are fabricated in DLSPPW system to realize the function of the three PDBSs in previous scheme, thus our approach further simplifies the design of the CNOT gate. The truth table and entangling ability of the gate have been investigated, and its good qualities have been demonstrated. The results show that such quantum plasmonic platform is quite promising for its ultra-high integration, low loss, and simple design, and will open a new way in future quantum information processing.

**Acknowledgments:** This work was supported by the National Key Projects for Basic Researches of China (No. 2012CB933501, No. 2012CB921802 and No. 2011CBA00205), the National Natural Science Foundation of China (Nos. 11322439, 11321063, 91321312).

**Figure legends:**

**Fig. 1. Characterization of PDBS.** (A) Schematic representation, with the inset showing the scheme of the simplified CNOT gate composed of three PDBSs proposed in Ref. [20]; (B) The SEM image of the configuration of the PDBS on the metal film. The input grating is composed of two gratings: the front gratings have five grooves with depth $d=30nm$, length $l=5\mu m$, width $w=200nm$, and period $p=540nm$; and the back gratings with different $w=100nm$ and $p=270nm$, aiming to obtain a directionally high conversion from spatial $p$-polarized light to SPP. The SPP reflecting grating comprises three grooves with $d=50nm$, $w=100nm$ and $p=380nm$. The output slits have $l=5\mu m$, $w=600nm$, and $p=830nm$. (C)-(E) show the CCD images of the output areas with $p$-polarized light input from left grating (C), $p$-polarized light input from right grating (D) and $s$-polarized light input from left grating (E), where the blue arrows remark the areas of input lights and their polarizations.

**Fig. 2. Experimental setup.** The experimental setup can be divided into three parts. The first part is the source: photon pairs at wavelength $\lambda=785nm$ are generated via spontaneous parametric down conversion (SPDC) in a *2mm* β-barium borate (BBO) crystal cut for type-II non-collinear phase matching, pumped by a *392.5nm* femtosecond laser with power $P=700mW$. A *10nm* interference filter (IF) with the center wavelength of *785nm* is used. Photons with perpendicular polarization states are separated by a polarizing beam-splitter (PBS) and coupled into single mode fibers (SMF). A delay line (DL) is inserted to control the temporal superposition of the photons. The polarization controller [PC1, quarter wave plate (QWP) + half wave plate (HWP) + QWP + Glan prism] are used to compensate polarization rotation in the SMFs. The second part is the microscopy system, composed of the CNOT gate sample (see inset), an input objective (×*50*, *N.A.=0.4*) and an oil objective (×*100*, *N.A.=1.32*). The final part represents the collection and analysis apparatus: the photons from two outputs are separated by a triangular reflector. After selected by PC2 (HWP + Glan prism), photons are then delivered to silicon avalanche photodiodes (APD) through multimode fibers (MMF) and coincidence measurements at the single photon counting modules (SPCM).

**Fig. 3. HOM interference results.** (A) HOM interference of the photon pair source; (B) HOM interference of SPP at the DLSPPW-based PDBS. Black dots correspond to measured results and red lines correspond to fitting curves.

**Fig. 4. Characterization of the integrated gate.** (A) Measured truth table for the gate; (B) Coincidence for non-orthogonal bases, with polarization controller set to pass $|s\rangle$ (black dots) and $|s\rangle+|p\rangle$ (red dots), and lines are fitting sine curves.

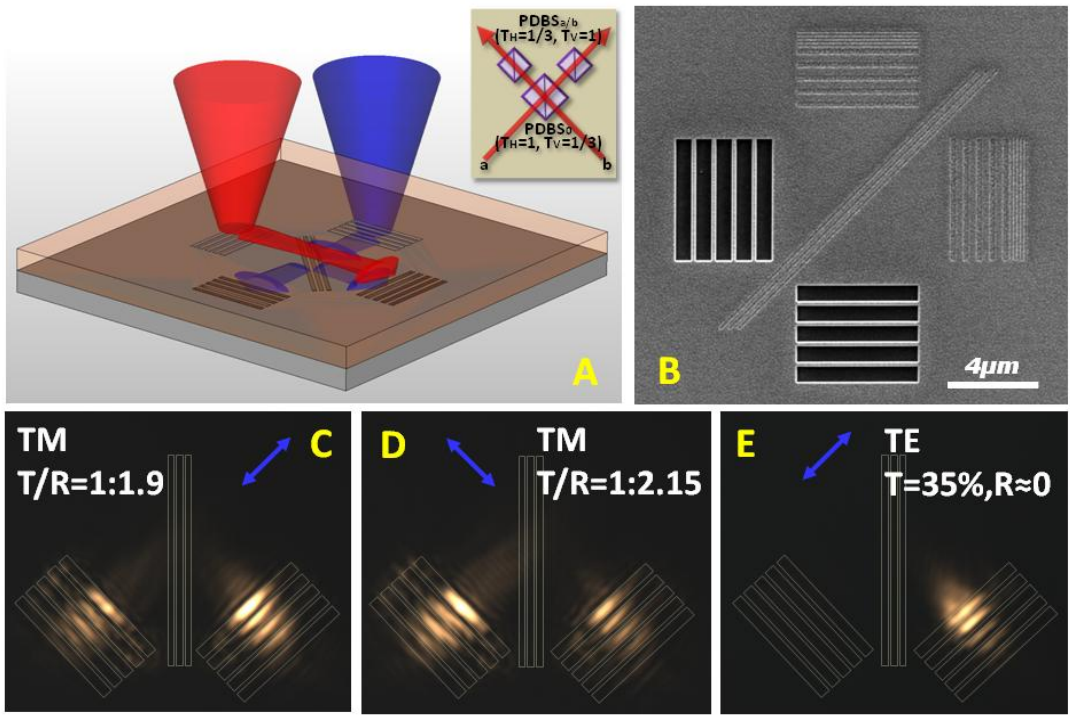

Fig.1

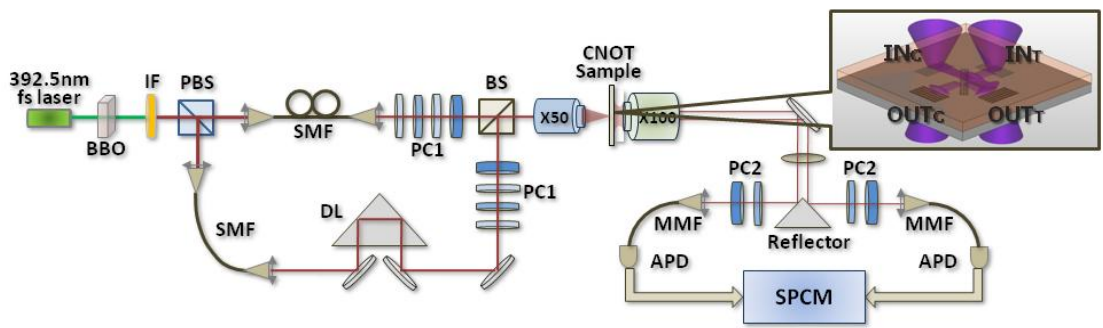

Fig.2

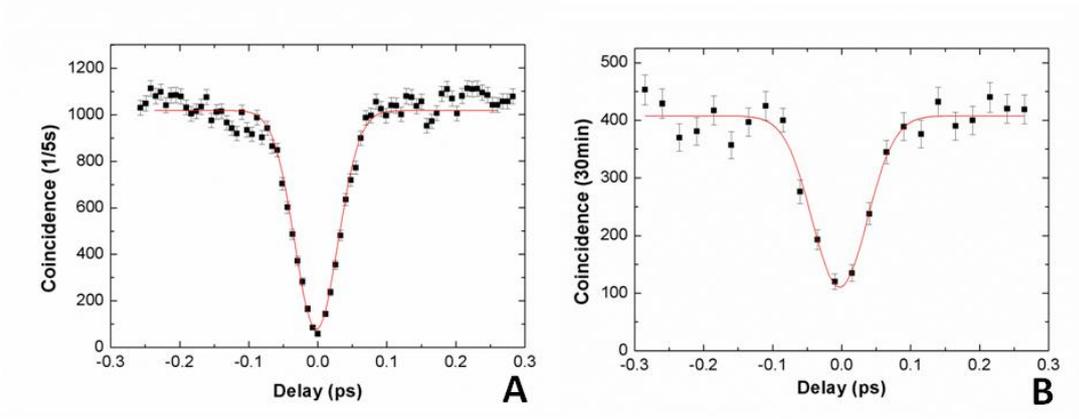

Fig.3

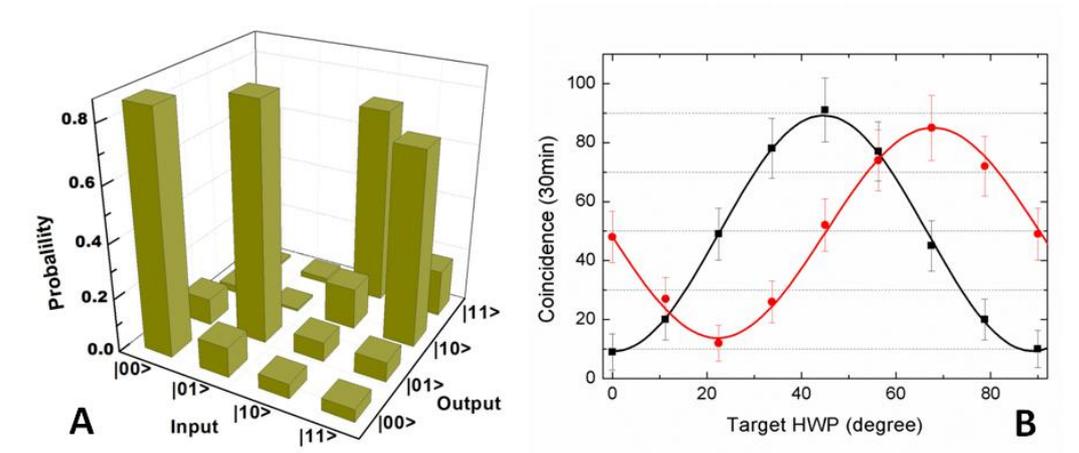

Fig.4

# Supplementary Materials for
# Integrated Quantum Controlled-NOT Gate Based on Dielectric-Loaded Surface Plasmon Polariton Waveguide

S. M. Wang[1,*,†], Q. Q. Cheng[2,*], Y. X. Gong[3], P. Xu[1], L. Li[1], T. Li[2,†], and S. N. Zhu[1,†]

## 1. Mode analysis of DLSPPW device

To guarantee only the lowest TE and TM (SPP) mode wave existing in the DLSPPW, the thickness of the loaded dielectric should be carefully chosen. We theoretically derive the dispersion relation of the TE and TM modes in the DLSPPW system dependent on the thickness of the loaded dielectric, **t**. The permittivity of silver is quoted from literature [P. B. Johnson and R. W. Christy, Phys. Rev. B **6**, 4370-4379 (1972)], and the index of silica is *1.5*. The dispersion relation of the system is shown in Fig. S1 A. The region in the green box in Fig. S1 A satisfies the requirement that only the lowest TE and TM mode simultaneously exist. We choose the thickness to be **t**=*250nm* pointed with the red dash dot line. The electric field distributions of the lowest TM (SPP) and TE mode wave are shown in Fig. S1 B and C. Their field locations are quite different which also enable the separately controlling of these two waves in the DLSPPW gate.

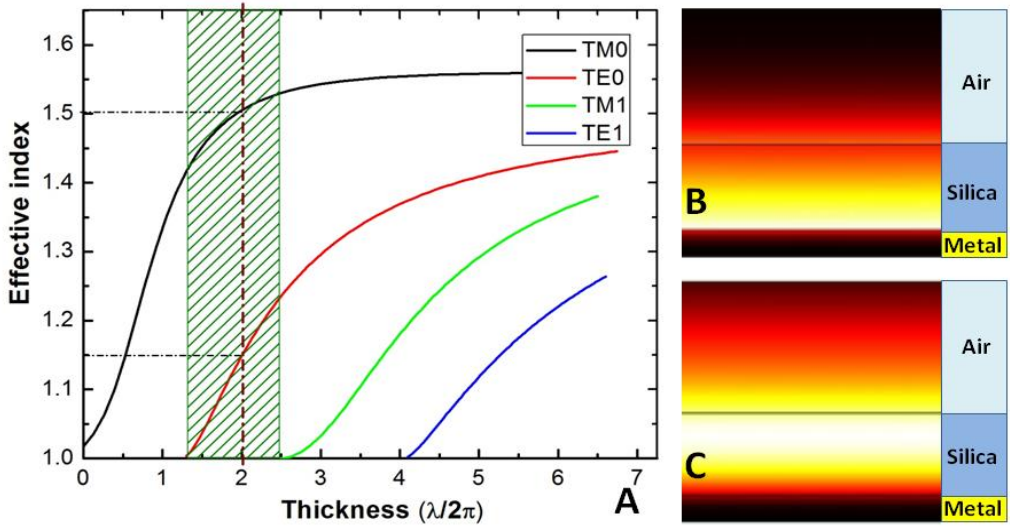

Fig. S1: (A) The dispersion relations of the waveguide modes in the DLSPPW. (B) and (C) show the electric field distributions of the lowest TM (TM0, SPP) and TE (TE0) mode wave, respectively.

## 2. TM reflector and TE transmitter

The SPP reflector in the DLSPPW gate is the crucial part of the device, whose parameters should be carefully chosen. Among a large number of samples, we choose the three-groove grating with width *w=100nm*, depth *d=50nm*, and period *p=380nm* to obtain the required *1: 2* T/R ratio. With different depths of the grooves, the T/R ratio of the SPP can be easily tuned. The outputs of the SPPs passing through the reflectors with different groove depths are presented in Fig. S2 A-D. For a deeper groove, the reflection of SPP will be larger, and the scattering loss at the reflector will also be increased, which is not desired for us. The width of the grooves has a less influence on the T/R ratio.

The output of TE mode wave can also be tuned by carefully choosing the parameters of the output slit. For required transmission of the TE mode wave, we choose the period of the output slit to be *830nm*. Figure S2 E and F give out the transmissions of the TE wave with different periods of output slits with **p**=*830nm* and *530nm*, respectively. It seems that the narrower slits have lower output, which is not required.

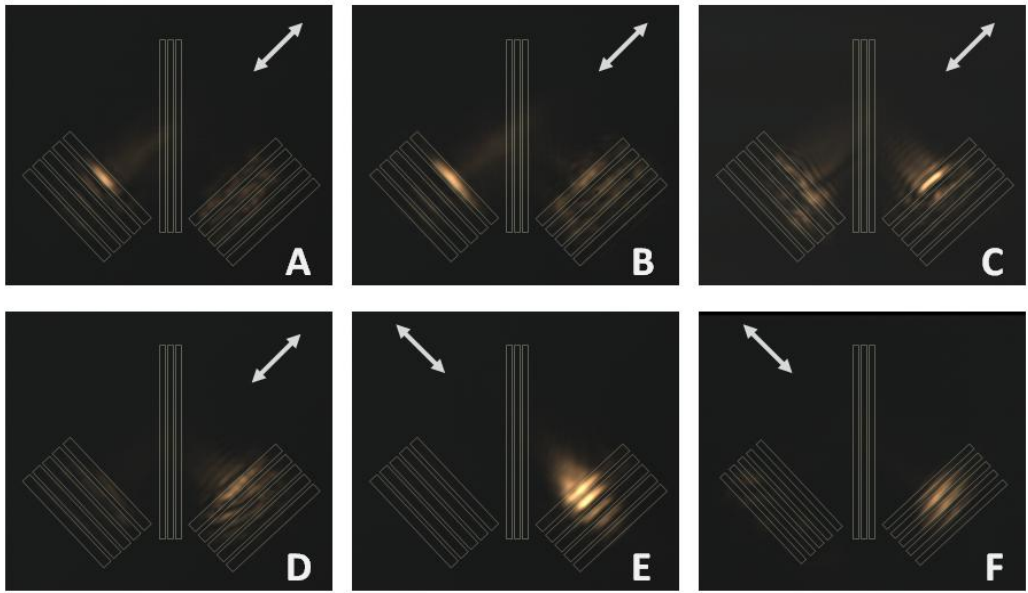

Fig.S2: (A)-(D) show the SPP outputs with different milling depths of the reflector groove, with **d**=*30nm*, *40nm*, *50nm*, and *60nm*, respectively. (E) and (F) show the TE wave outputs with different period of the output slit, with **p**=*830nm* and *530nm*. The white arrows remark the areas of input lights and their polarizations.

## 3. Phase compensation and decoherence

Since the effective index of the TM (SPP) and TE mode wave in the DLSPPW system are different (*1.5* and *1.15*), the phase difference between these two polarized waves can be easily compensated by a phase compensator. The *45°* input can be collected by CCD camera with PC2 setting to *45°* pass, with the setup being shown in Fig. S3 A. To see the purity of the output light, the PC2 is set to *45°* and *-45°* pass. By using a well placed phase compensator, the output power ratio between *45°/-45°* polarizations is about *10:1*, which is good enough for the CNOT gate (Fig. S3 B and C).

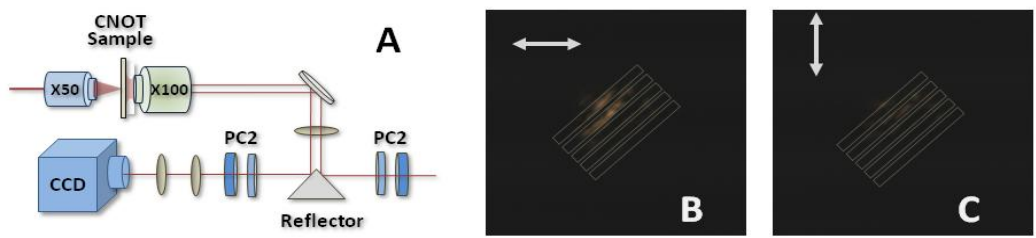

Fig. S3: (A) The schematic representation of the collection part of the setup, with one output field being observed by a CCD camera. (B) and (C) show the CCD image of the output for s+p-polarized (*45 º* oblique to the output slits) polarization input and *45 º /-45 º* collection, respectively.

It should also be mentioned that the phase-damping decoherence of the photons generated from parametric down conversion will occur in such DLSPPW devices. However, we can see that the phase difference, about *4λ*, is much smaller than the coherence length of the photon pair source, which is about *100λ*, see Fig. 3 A. Therefore, we can conclude that the high birefringence devices, such as the DLSPPW system, is also capable to be used in polarization-encoded quantum information processing, as long as the phase difference is smaller than the coherence length.